\documentclass[letterpaper]{article} 
\usepackage[submission]{aaai24}  
\usepackage{times}  
\usepackage{helvet}  
\usepackage{courier}  
\usepackage[hyphens]{url}  
\usepackage{graphicx} 
\def\UrlFont{\rm}  
\usepackage{natbib}  
\usepackage{caption} 
\usepackage{algorithm}
\usepackage{algorithmic}

\usepackage{newfloat}
\usepackage{listings}
\DeclareCaptionStyle{ruled}{labelfont=normalfont,labelsep=colon,strut=off} 
\floatstyle{ruled}
\newfloat{listing}{tb}{lst}{}
\floatname{listing}{Listing}
\usepackage{soul}
\usepackage{url}
\usepackage{graphicx}
\usepackage{amsmath}
 
\usepackage{amsthm}
\usepackage{booktabs}
\usepackage{algorithm}
\usepackage{algorithmic}
\usepackage[switch]{lineno}
\usepackage{ragged2e} 
\usepackage{booktabs,makecell, multirow, tabularx}
\usepackage{multicol}
\usepackage{amssymb}
\usepackage{bbding}

\title{DELTA: Dynamic Embedding Learning with Truncated\\
Conscious Attention for CTR Prediction}
\def\showauthors@on{Yes}
\author {
    Chen Zhu\textsuperscript{\rm 1},
    Liang Du\textsuperscript{\rm 2},
    Hong Chen\textsuperscript{\rm 1},
    Shuang Zhao\textsuperscript{\rm 2},
    Zixun Sun\textsuperscript{\rm 2},
    Xin Wang\textsuperscript{\rm 1}\equalcontrib,
    Wenwu Zhu\textsuperscript{\rm 1}\equalcontrib
}
\affiliations {
    \textsuperscript{\rm 1}Tsinghua University\\
    \textsuperscript{\rm 2}Tencent\\
    \{zhu-c21,chen20\}@mails.tsinghua.edu.cn, \{xin\_wang,wwzhu\}@tsinghua.edu.cn,\\
    \{lucasdu,nicholezhao,zixunsun\}@tencent.com
}

\begin{document}
\maketitle


\begin{abstract}
Click-Through Rate (CTR) prediction is a pivotal task in product and content recommendation, where learning effective feature embeddings is of great significance.
However, traditional methods typically learn fixed feature representations without dynamically refining feature representations according to the context information, leading to suboptimal performance. Some recent approaches attempt to address this issue by learning bit-wise weights or augmented embeddings for feature representations, but suffer from uninformative or redundant features in the context. 
To tackle this problem, inspired by the Global Workspace Theory in conscious processing, which posits that only a specific subset of the product features are pertinent while the rest can be noisy and even detrimental to human-click behaviors,
we propose a CTR model that enables \textbf{D}ynamic \textbf{E}mbedding \textbf{L}earning  with \textbf{T}runcated Conscious \textbf{A}ttention for CTR prediction, termed DELTA. DELTA contains two key components:
(I) conscious truncation module (CTM), which utilizes curriculum learning to apply adaptive truncation on attention weights to select the most critical feature in the context;
(II) explicit embedding optimization (EEO), which applies an auxiliary task during training that directly and independently propagates the gradient from the loss layer to the embedding layer, thereby optimizing the embedding explicitly via linear feature crossing.
Extensive experiments on five challenging CTR datasets demonstrate that DELTA achieves new state-of-art performance among current CTR methods.
Codes and models will be released at \url{https://github.com/DELTA-C4}.
\end{abstract}

\section{Introduction}
The prediction of Click-Through Rate (CTR) is a critical task in online advertising~\cite{wang2021dcn,shan2016deep} and recommender systems~\cite{zhou2019deep}. 
Accurate predictions of the CTR not only drive revenue for online platforms but also enhance the user experience by presenting relevant content. Many models have been proposed for CTR, such as Logistic Regression (LR)~\cite{richardson2007predicting}, POLY2~\cite{chang2010training}, and tree-based methods~\cite{he2014practical}. In recent years, employing feature embeddings\cite{rendle2010fm,song2019autoint,lian2018xdeepfm} has become a common means to augment the model's representational capacity, which can effectively capture the intricate relationships and patterns within and between the data~\cite{zhang2021deep}.

\begin{figure}[t]
  \centering
   \includegraphics[width=0.7\linewidth]{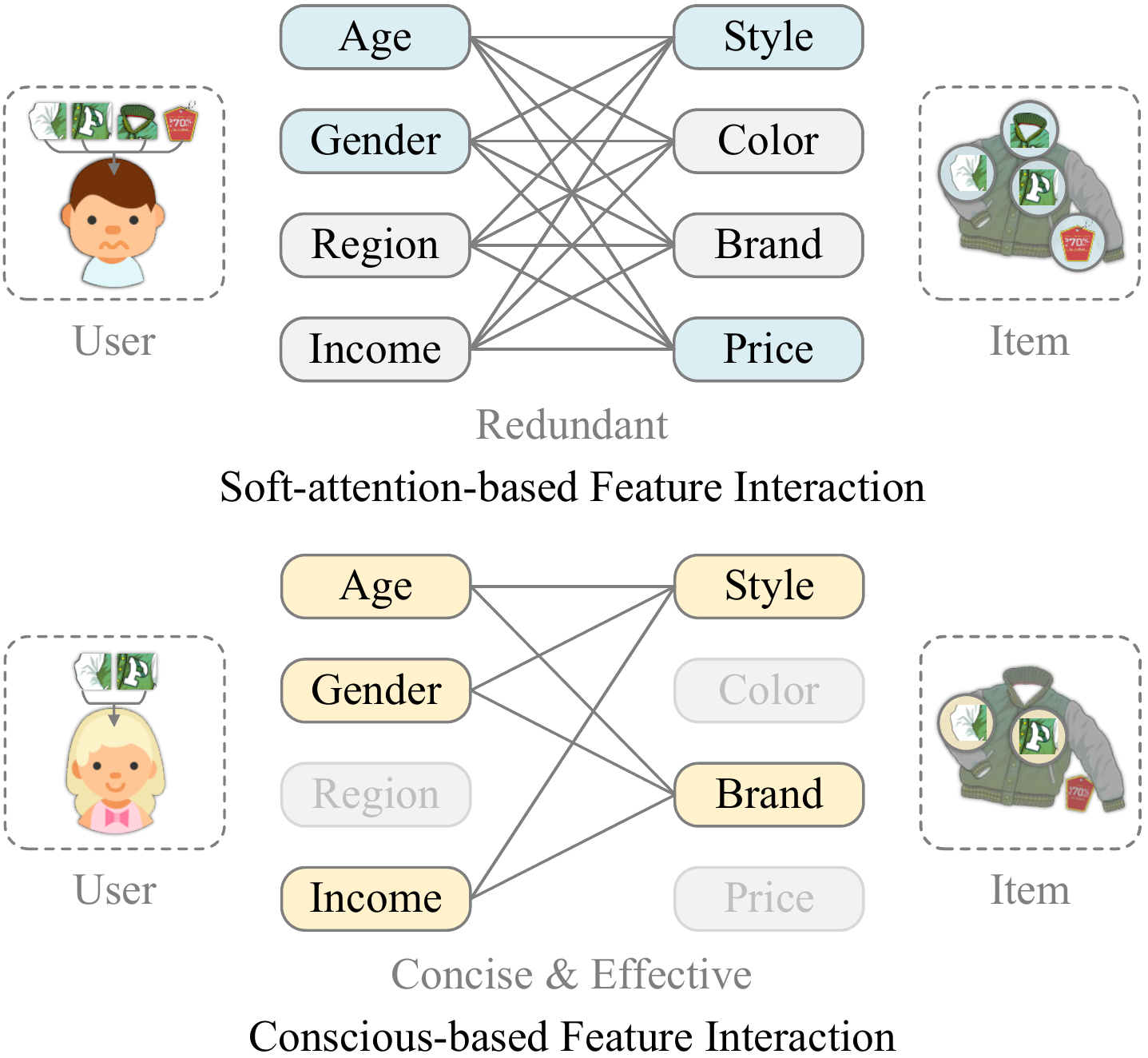}
  \caption{
  %
  %
Illustration of the traditional soft-attention-based feature interaction and the proposed conscious feature interaction.
%
}
  \label{fig0}
  
\end{figure}



However, most existing methods typically learn fixed feature embeddings for each feature field~\cite{rendle2010fm,guo2017deepfm}, which lack the flexibility to adapt to varying context information. Some recent approaches have attempted to address this issue by assigning each feature with multiple embedding vectors~\cite{lian2018xdeepfm,yang2020operation}, each corresponding to a specific field. However, they still essentially learn fixed embeddings as these vectors do not adapt to changes in context information. Recently, more sophisticated methods such as MaskNet~\cite{wang2021masknet} and Frnet~\cite{wang2022frnet} have been proposed to learn dynamic embeddings. MaskNet introduces instance-guided feature-wise masks to highlight essential elements, while Frnet combines soft attention and Multi-Layer Perceptron (MLP) to learn context-aware embeddings. 

Despite these advancements, all of the aforementioned methods include all features during context understanding and embedding refining, neglecting the noise introduced by redundant features which may do harm to the CTR prediction, leading to inferior performance and higher computation complexity. As shown in Figure~\ref{fig0}, different users will focus on different specific features. For example, assuming that the female is a wealthy person. When choosing clothes, she will only take ``Style'' and ``Brand'' of the cloth into consideration, while other information such as ``Price'' will not be involved in the thinking process. Additionally, according to ``conscious processing~\cite{baars2005global}'', the human brain directly ignores this redundant information while existing methods still take this into consideration.
Besides, previous works like DCN~\cite{wang2017deep} and xDeepFM~\cite{juan2016field} have proved that combining both explicit linear feature interactions and implicit non-linear feature interactions for prediction is more effective than only using the implicit interactions. 
%
However, existing dynamic embedding methods typically rely on MLPs to generate final predictions~\cite{wang2021masknet,mao2023finalmlp}, or aggregate linear and deep semantic non-linear features to obtain richer feature representationss~\cite{wang2022frnet}, without fully leveraging the interplay between linear and non-linear feature interactions. However, simply adding a linear branch in these models will result in embedding representation degradation and global performance decrease, since the sophisticated non-linear feature extraction process will be affected by the coarse linear features that are combined together,

To tackle these challenges, we propose a novel model, namely DELTA, which enables \textbf{D}ynamic \textbf{E}mbedding \textbf{L}earning  with \textbf{T}runcated Conscious \textbf{A}ttention utilizing \textbf{conscious truncation module (CTM)} and explicit embedding optimization (EEO). More
specifically, we draw inspiration from the Global Workspace Theory (GWT) in conscious processing~\cite{baars2005global} that ``conscious attention'' focuses on a limited number of essential elements and is believed to contribute to humans' rapid decision-making and efficient learning. This property makes conscious attention superior as weights for irrelevant features in vanilla soft attention are never 0. To mimic human conscious processing, we proposed CTM, which learns and conducts dynamic truncation on attention weights, thus generating a bottleneck structure that limits the features that attention can attend to and reduces computation complexity. 
Moreover, we rethink the combination issue of linear and non-linear branches and leverage \textbf{explicit embedding optimization (EEO)} to learn linear representation, which is independent of the non-linear branch.
As an explicit feature interaction branch, the EEO performs linear feature crossing like previous methods, while the extracted feature is not merged to the neural network-based implicit branch for the final prediction. It directly propagates the gradient from the loss layer to the embedding layer to enhance the crucial embeddings for further feature combinations. 
%

In summary, the proposed CTM and EEO aim to select the most important features and to learn dynamic embedding with respect to the context information for the CTR prediction in an explicit way. The main contributions of this paper are summarized as follows:
\begin{itemize}
    \item We propose a CTR model that mimics human conscious processing to fundamentally boost performance.
    \item A conscious truncation module (CTM) is introduced that leverages curriculum learning strategy to apply adaptive 
    truncated conscious attention (semi-hard) to select the most critical feature in different contexts, which boosts the performance as well as reduces the computational complexity.
    \item Explicit embedding optimization (EEO) is proposed that directly and independently propagates loss gradient to the embedding layer to enhance the crucial embeddings via explicit feature crossing, which requires no extra cost during inference.
    \item Extensive experiments on Criteo, Avazu, Malware, Frappe, and MovieLens datasets demonstrate that our method achieves new state-of-the-art performance.
\end{itemize}

\section{Related works}
\begin{figure*}[htb]
  \centering
  \includegraphics[width=1\linewidth]{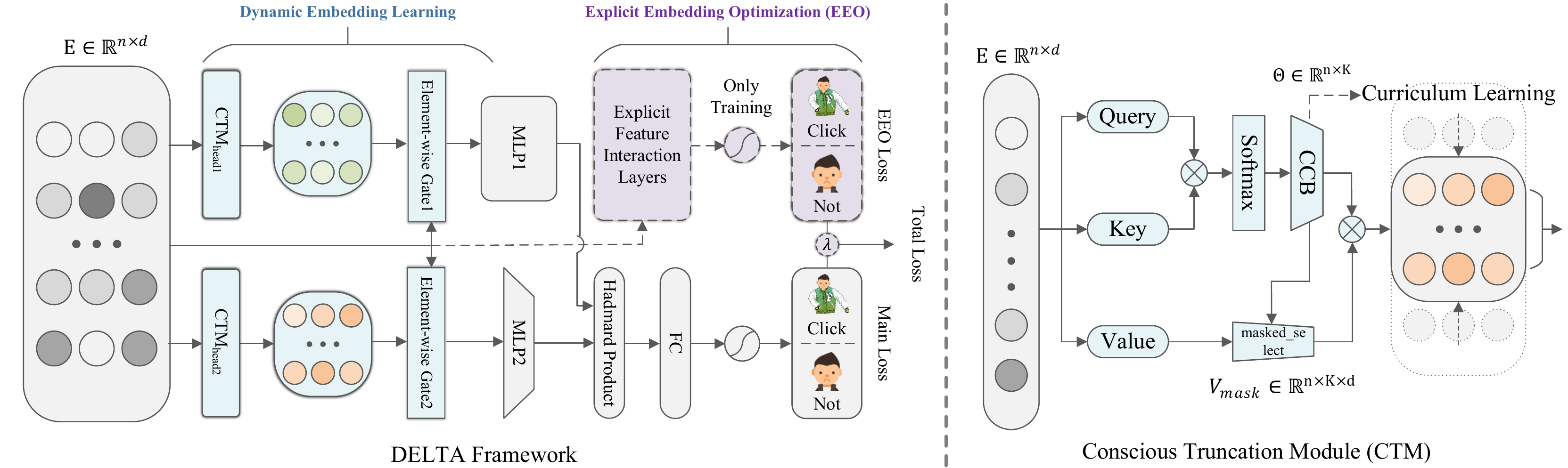}
  \caption{
  Overview framework of the proposed DELTA. CCB is the abbreviation of the Curriculum Conscious Bottleneck.
  }
\label{fig1}
\end{figure*}

\noindent\textbf{Feature Interaction Learning }
Many CTR prediction methods have achieved huge success by
modeling feature interactions to enrich feature representations.
LR~\cite{richardson2007predicting} explicitly models first-order interactions while POLY2~\cite{chang2010training} utilizes a degree-2 polynomial mapping to capture second-order interactions but bring higher computational complexity. To address this issue, FM~\cite{rendle2010fm} uses latent vectors to learn the pairwise feature interactions explicitly. As extensions of FM, FFM~\cite{juan2016field} and FwFM~\cite{pan2018field} take into account the field a feature belongs to in order to further improve performance. However, these methods are unable to capture high-order interactions. To solve this problem, most CTR models utilize deep neural networks to model complex feature interactions implicitly~\cite{zheng2022implicit}. Wide\&Deep~\cite{cheng2016wide} combines a DNN with LR to improve the model's memorization and generalization abilities. DeepFM~\cite{guo2017deepfm} replaces the LR part with FM to alleviate empirical feature engineering in feature engineering. Some models also use a sub-network to learn high-order interactions~\cite{zheng2022hien,zhang2022deep,shen2022hierarchically}. DCN~\cite{wang2017deep} replaces the LR part in Wide\&Deep with cross-net with residue connection to learn cross features. Autoint~\cite{song2019autoint} utilizes the multi-head self-attention mechanism to explicitly model the feature interactions in the low-dimensional space. InterHAt~\cite{li2020interpretable} and DESTINE~\cite{xu2021disentangled} further enhance the performance by employing hierarchical attentions and disentangling the unary part in self-attention. 

Nevertheless, it is important to note that all of the aforementioned methods only learn a fixed embedding/embeddings for feature interactions without considering the varying significance of each feature under different contexts, which consequently leads to suboptimal performance.\\

\noindent\textbf{Feature Embedding Learning }
In recent years, some methods have been proposed to learn dynamic embeddings instead of fixed embeddings. FFM~\cite{juan2016field} learns field-aware embeddings but still neglects the context information. FiBiNET~\cite{huang2019fibinet} makes a step forward by employing the Squeeze-and-Excitation network to apply a vector-wise reweighing procedure to the original features. To facilitate more fine-grained learning of embeddings,
ContextNet~\cite{wang2021contextnet} proposed instance-guided bit-wise masks to highlight the essential elements while MaskNet~\cite{wang2021masknet} further expands this method by employing sequential or parallel masks. To further extend, FinalMLP applies two distinct masks generated from different empirically selected feature sets and Frnet~\cite{wang2022frnet} utilizes self-attention and MLP to learn context-aware embeddings.  
\begin{table}[thb]
 \small
 \centering
  \begin{tabular}{c|c|c|c|c}
  \toprule
{Model} & {Granularity} & Selection
 & Learning & Fusion\\
\midrule
    FiBiNET  & Vector & \XSolidBrush & Imp & $\odot$ \\
    ContextNet & Bit & \XSolidBrush&Imp&$\odot$ \\
    MaskNet & Bit & \XSolidBrush &Imp& $\odot$ \\
    FRNet & Bit & \XSolidBrush&Imp\&Exp& +\\
    FinalMlp & Bit & \Checkmark(manual)&Imp& $\odot$\\
\midrule
    \textbf{DELTA} & Bit &\Checkmark(auto)&Exp& +\\

\bottomrule
  \end{tabular}
   \caption{The connection and difference between DELTA and existing context-aware embedding learning models. Selection means whether the model processes the context selectively. Learning means the context-aware embeddings are learned implicitly(Imp) or explicitly(Exp). Fusion means how the model fusion context-aware embeddings and original embeddings, $\odot$ stands for Hadamard product and + stands for weighted sum}
 \label{tab_compare}
\end{table}

As delineated in Table~\ref{tab_compare}, although our method shares certain aspects with existing methods, it is fundamentally different at its core. While all of the above methods utilize MLP to learn context-aware embeddings implicitly and do not filter the context automatically, our proposed DELTA mimics conscious processing to solve the common feature redundancy problem ignored by previous mask-based and attention-based methods and learns context-aware embeddings explicitly.\\

\noindent\textbf{Curriculum Learning }
Curriculum learning is an approach to machine learning that uses a curriculum of tasks to train a model in a sequence that gradually increases in difficulty~\cite{bengio2009curriculum}. This approach enables machines to learn more structured and efficiently, allowing them to focus on simpler tasks before progressing to more complex tasks~\cite{wang2021survey,chen2021curriculum}. The key problem in curriculum learning is designing easy-to-hard curricula, and the difficulty lies in defining the degree of difficulty of the curriculum. 

\section{Method}

\subsection{Overview framework}
In this section, we present the overview framework of our proposed DELTA, which consists of several key components as illustrated in Figure~\ref{fig1}. The computation process of CTM, EFG, and EEO will be illustrated in the following subsections.\\

\noindent\textbf{Embedding layer.}
%
The embedding layer maps each high-dimensional raw feature to a dense, continuous vector of size $d$. Specifically, let $x_1, x_2, ..., x_n$ be the one-hot representation of categorical variables, and let $V$ be the vocabulary size. The embedding of $x_i$ can be calculated as:
\begin{equation}\label{eq1}
\mathbf{e}_i = \mathbf{E_i} x_i
\end{equation}
where $\mathbf{E_i} \in \mathbb{R}^{V \times d}$ is the embedding matrix and $\mathbf{e}_i \in \mathbb{R}^d$ is the embedding vector for $x_i$. \\

\noindent\textbf{DELTA network.} After the user and item features are represented as dense embeddings, they are passed to our proposed DELTA network. 
Taking inspiration from high-order thought(HOT) theory in consciousness\cite{byrne1997some}, which postulates that consciousness consists of low-order mental states and high-order thoughts, we employ CTM with two heads. The first head is followed by a deep and narrow MLP to simulate high-order thoughts while the second employs a shallow and wide MLP to simulate low-order mental states. The computation process of our DELTA network can be summarized as follows:
\begin{enumerate}
    \item Embeddings are concatenated and then passed to $\mathrm{CTM}_{\mathrm{head1}}$ and $\mathrm{CTM}_{\mathrm{head2}}$. 
    \item $\mathrm{CTM}_{\mathrm{head1}}$ and $\mathrm{CTM}_{\mathrm{head2}}$ generate context-aware enhanced embeddings.  Note that we don't reduce the dimensionality during linear projection of each head for further fusion.
    \item The Element-wise Fusion Gate(EFG) fusions original embeddings and enhanced embeddings to obtain dynamic embeddings.
    \item Dynamic embeddings are then passed to MLP1 and MLP2 for further interactions.
    \item Then, the final fully-connected layer outputs the prediction $\hat{y}_{\rm{main}}$ using a sigmoid function.
    \item As an auxiliary task, the original embeddings $\mathbf{E}$ are passed to EEO for explicit interactions. Note that the network outputs the prediction $\hat{y}_{\rm{EEO}}$ apart from $\hat{y}_{\rm{main}}$. 
\end{enumerate}

%

\noindent\textbf{Loss function.}
We use the widely used binary cross-entropy loss to train DELTA. This loss function is defined as:
\begin{equation}\label{eq3}
 L = -\frac{1}{N}\sum_{i=1}^{N} y_i \log(\hat{y}_i) + (1 - y_i) \log(1 - \hat{y}_i),
\end{equation}
where $N$ is the number of examples, $y_i$ is the true label, and $\hat{y}_i$ is the predicted probability of a click. During the training, we use the weighted sum of two log loss calculated from the two probabilities to conduct gradient descent:
\begin{equation}\label{eq4}
L_{\rm{total}} = L_{\rm{main}}+\lambda L_{\rm{EEO}}, 
\end{equation}
where $\lambda$ is the loss weight of the EEO. Note that DELTA utilizes $L_{\rm{total}}$ to perform gradient descent during training while using $\hat{y}_{\rm{main}}$ for inference, and the rationale will be discussed in the EEO subsection.

\subsection{Conscious truncation module (CTM)}
\label{module1}
To fully understand and utilize the information under different contexts to learn dynamic embeddings, we proposed a novel module named \textbf{C}onscious \textbf{T}runcation \textbf{M}odule (CTM). The word ``consciousness'' is widely used in different senses. In this paper, we consider ``consciousness'' as the ``Global availability'' introduced by~\cite{dehaene2017consciousness}, which corresponds to the transitive meaning of consciousness (as in \textit{``The user is conscious of the color of the item''}). When the user interacts with the item, only a few features can be attended to among the vast features available, while the rest remain unconscious~\cite{zhao2021consciousness}. Conscious thought is a set of these features we have become aware of, joined together, and made globally available to others~\cite {bengio2017consciousness}. \\

\noindent\textbf{CTM structure.} Inspired by the mechanism of human conscious processing, we propose a novel truncation that simulates the conscious processing of the users. Instead of the soft-attention, which uses a universal weighted average as the output, we use the highest top-k attention weighted average as the output, which helps the model focus on the most relevant features while ignoring the influence of irrelevant and noisy features.

Figure~\ref{fig1} shows the structure of the CTM. The computation process of the CTM unit can be formulated as follows:

First, let the input to the self-attention mechanism be $X \in \mathbb{R}^{n \times d}$, where $n$ is the number of feature fields, and $d$ is the dimensionality of the feature embedding. We define the query, key, and value matrices as:
\begin{equation}\label{eq5}
 Q = XW_Q, K = XW_K, V = XW_V, 
\end{equation}
where $W_Q, W_K, W_V \in \mathbb{R}^{d \times d}$ are the weight matrices for the query, key, and value, respectively.

Next, compute the similarity between the query and key matrices using the dot product and divide it by the square root of the key dimensionality. Then, we apply the softmax function to this matrix to obtain the attention weights $w$:
\begin{equation}\label{eq6}
w = softmax(\frac{QK^T}{\sqrt{d_k}}). 
\end{equation}
After softmax, we calculate consciousness-inspired truncation based on attention weights:
\begin{equation}\label{eq7}
\theta_i =  \begin{cases} w_i &\text{ if $w_i \ge w_{top-k}$ }\\ \emptyset &\text{ otherwise} \end{cases},
\end{equation}
where $k$ is the size of the consciousness bottleneck and $\sigma_{top-k}$ is the k-th highest weight in attention weights.

Finally, we use truncated attention weights to compute the weighted sum of the value matrix as $\theta\cdot V_{mask}$.
This weighted sum of the values is then flattened and deemed as enhanced embeddings $\mathbb{E}\in \mathbb{R}^{1 \times n\cdot d}$.\\

\noindent\textbf{Dynamic consciousness bottleneck with curriculum learning.} For the proposed CTM, the size of the consciousness bottleneck should be dynamic to select the most critical feature under different contexts adaptively. Unlike other hyper-parameters, the bottleneck size can reflect the task's difficulty, big bottlenecks allow all information to pass and represent easy tasks, while small bottlenecks limit the information passed and can be seen as hard tasks. Therefore, we utilize curriculum learning to design easy to hard curricula using Algorithm~\ref{alg1},
 which enables our model to build on its previous knowledge and improve its overall performance. 

In Algorithm~\ref{alg1}, 
$Flag$ represents Whether the bottleneck has decreased at the current learning rate, and $\Delta$ stands for bottleneck decrease speed when the model converges at current curricula. $\delta$ is the learning rate reduction rate, $\mathcal{C}_{max}$ is the number of feature fields, and $t_{max}$ is the maximum training epoch.\\
\begin{algorithm}[htp]
    \caption{Curriculum learning algorithm for CTM}
    \label{alg1}
    \begin{algorithmic}[1]
    \STATE \textbf{Input}: $\mathcal{R}_t$ - Learning rate at t-th epoch, $\mathcal{L}_t$ - Logloss at t-th epoch, $\mathcal{C}_t$ - Consciousness bottleneck size at t-th epoch, $\mathcal{M}$ - DELTA model, $D$ - dataset\\
    \STATE \textbf{Parameter}: $Flag$, $\Delta$, $\delta$,  $\mathcal{C}_{max}$, $t_{max}$\\
    \STATE \textbf{Output:} $\mathcal{C}_{t+1}$,$\mathcal{R}_{t+1}$
    \STATE Initialize $Flag \leftarrow 0$ and $\mathcal{C}_{0}\leftarrow\mathcal{C}_{max}$
    \WHILE{$t \leq t_{max}$}
        \STATE Train $\mathcal{M}$ on D
        \STATE Update $\mathcal{M}$ with the new parameters
        \STATE Evaluate $\mathcal{M}$ on the validation set
        \STATE Update $\mathcal{L}_{t+1}$
        \IF{$\mathcal{L}_{t+1}$ \textgreater $\mathcal{L}_{t}$}
        \IF{not $Flag$}
        \STATE $\mathcal{C}_{t+1}=\mathcal{C}_{t}-\Delta$
        \STATE Set $Flag=1$      
        \ELSE
        \STATE $\mathcal{R}_{t+1}=\delta\cdot\mathcal{R}_{t}$
        \STATE Set $Flag=0$  
        \ENDIF
        \ELSE
        \STATE Set $Flag=0$         
        \ENDIF
    \ENDWHILE
        \STATE \textbf{return} $\mathcal{C}_{t+1}$,$\mathcal{R}_{t+1}$
    \end{algorithmic}
\end{algorithm} 

\noindent\textbf{Computation complexity analysis.} Directly employing self-attention to embeddings and conducting full interaction results in a complexity of $O(d^2nh)$, where $h$ is the number of heads and $h$ equals 2 in DELTA, while the bottleneck lowers it to $O(dKnh)$, where $K$ is the size of the information bottleneck and $K<d$.\\

\noindent\textbf{Discussion.}
Our ability to make decisions quickly across different items is attributed to the computation involved in ``human conscious processing'', which is introduced in GWT~\cite{baars2005global,vanrullen2021deep} and explained in Yoshua Bengio's recent work and other cognitive science researches~\cite{zhao2021consciousness,koch2007attention}. A central characterization of conscious attention is that it involves a bottleneck, which forces one to handle dependencies between very few environmental features at a time. 
In vanilla soft attention~\cite{vaswani2017attention}, weights for irrelevant features are never 0, and learning vital features will be more difficult.
\subsection{Element-wise fusion gate (EFG)}
While our proposed CTM can effectively capture the pair-wise feature significance in different contexts, it is also necessary to model the general influence of each feature field~\cite{xu2021disentangled}. We apply an element-wise gate $gate \in \mathbb{R}^{1 \times n\cdot d} $ containing $n\cdot d$ learnable parameters to fusion the original embeddings and the enhanced embeddings. We apply a sigmoid function of the gate to limit the weight of each element between [0,1], formulated as 
\begin{equation}\label{eq9}
\begin{aligned}
EFG1(\mathbf{E},\mathbb{E}_1)=\sigma(gate1)\times \mathbf{E} + (1-\sigma(gate1))\times\mathbb{E}_1,\\
EFG2(\mathbf{E},\mathbb{E}_2)=\sigma(gate2)\times \mathbf{E} + (1-\sigma(gate2))\times\mathbb{E}_2.
\end{aligned}
\end{equation}

The EFG not only integrates unary feature significance lies in original embeddings and binary feature relationships modeled by CTM, but also lets different heads of CTM focus on different aspects of the prediction task by applying two distinct gates, thus generating dynamic embeddings with better capability.

\subsection{Explicit embedding optimization (EEO)}\label{LFE}

Previous works like DeepFM~\cite{guo2017deepfm} perform linear feature interactions, and the linear feature is concatenated with the non-linear feature extracted by the MLP before decoding. This fusion can be formulated as 
\begin{equation}\label{eq10}
\hat{p}=I\cdot M=\sigma([I_{MLP},I_{linear}][W_{MLP},W_{linear}]),
\end{equation}
where $I$ and $W$ are the input and the weight matrix of the final fully connected layer. We reformulate this problem as generating predictions from aggerating the output of MLP and linear interaction branch:
\begin{equation}\label{eq11}
\hat{p}=\sigma(O_{MLP}+O_{linear})
\end{equation}
where $O_{MLP}=I_{MLP}\cdot W_{MLP}$ and $O_{linear}=I_{linear}\cdot W_{linear}$. However, the weight of $O_{linear}$ is fixed in Eq \ref{eq11} and as illustrated in~\cite{mao2023finalmlp}, MLP shows better performance than linear interaction and assigning the same weights to these two branches is inadequate. Nevertheless, it also introduces the mutual interference issue caused by the combination of features at different levels, which affects the prediction performance~\cite{bian2022can}.
%

%

Therefore, to fully exploit the important representations brought by explicit feature interaction, we propose explicit embedding optimization (EEO), disentangling the $O_{linear}$, deeming it as an auxiliary task, and assigning weight $\lambda$ to it. We disable EEO during inference since the sophisticated non-linear feature interactions outperform the coarse linear interactions.
Different from previous methods, the EEO independently models the high-order feature interactions, which directly propagates gradient from the loss layer to the embedding layer to enhance the crucial embeddings via linear feature interactions.  The EEO not only avoids the mutual interference from the combination of linear and non-linear features but also exploits the network's ability to adaptively enhance the specific crucial embeddings for further feature interaction. 
%
Our EEO can employ various explicit high-order architectures, in this paper, we adopt cross-net to instantiate the EEO.

\section{Experiments}
We conduct extensive experiments to answer the following questions:
\begin{itemize}
    \item \textbf{Q1:} How does DELTA perform compared to state-of-the-art methods for CTR prediction?
    \item \textbf{Q2:} What is the performance improvement of each proposed module compared with the baseline model? 
    \item\textbf{Q3:} How do the key hyper-parameters of DELTA (e.g., the bottleneck size of the CTM and the loss weight of the EEO) impact its performance? 
\end{itemize}

\subsection{Experimental Settings}

\begin{table*}[t]

 \small
 \centering
 \begin{tabular}{c|c|cc|cc|cc|cc|cc}
  \toprule
  \multirow{2} * {\makebox[0.01\textwidth][c]{Type}} &
  \multirow{2} * {\makebox[0.02\textwidth][c]{Method}} & 
  \multicolumn{2}{c|}{\makebox[0.005\textwidth][c]{Criteo}} & 
  \multicolumn{2}{c|}{\makebox[0.005\textwidth][c]{Avazu}} & 
  \multicolumn{2}{c|}{\makebox[0.005\textwidth][c]{Malware}} &
  \multicolumn{2}{c|}{\makebox[0.005\textwidth][c]{Frappe}} & 
  \multicolumn{2}{c}{\makebox[0.005\textwidth][c]{MovieLens}}  \\
  &       & AUC $\uparrow$  & Lloss $\downarrow$ & AUC $\uparrow$ & Lloss $\downarrow$ & AUC $\uparrow$  & Lloss $\downarrow$     & AUC $\uparrow$          & Lloss $\downarrow$     & AUC $\uparrow$          & Lloss $\downarrow$   \\
\midrule
  \multirow{1}*{S}
  & FM & 0.8028 & 0.4514 & 0.7720 & 0.3844 & 0.7309 & 0.6052 & 0.9708 & 0.1934 & 0.9391 & 0.2856 \\
\midrule
  \multirow{2}*{H}
  & NFM& 0.8072 & 0.4444 & 0.7811 & 0.3810 & 0.7352 & 0.5988 & 0.9746 & 0.1915 & 0.9437 & 0.2945 \\
  & OPNN & 0.8096 & 0.4426 & 0.7821 & 0.3829 & 0.7408 & 0.5840 & 0.9795 & 0.1805 & 0.9497 & 0.2704 \\
  & CIN & 0.8086 & 0.4437 & 0.7843 & 0.3783 & 0.7395 & 0.5967 & 0.9776 & 0.2010 & 0.9483 & 0.2808 \\
\midrule
  \multirow{8}*{E}
  & DCN   & 0.8106 & 0.4414 & 0.7853 & 0.3790  & 0.7403 & 0.5944 & 0.9789 & 0.1814 & 0.9458 & 0.2685 \\
  & DeepFM  & 0.8130 & 0.4389 & 0.7856 & 0.3794 & 0.7432 & 0.5924 & 0.9789 & 0.1770 & 0.9556 & 0.2497 \\
  & xDeepFM & 0.8127 & 0.4392 & 0.7851 & 0.3776 & 0.7435 & 0.5920 & 0.9792 & 0.1889 & 0.9578 & 0.2480 \\
  & AutoInt+  & 0.8128 & 0.4396 & 0.7852& 0.3769 & 0.7409 & 0.5939 & 0.9786 & 0.1890 & 0.9501 & 0.2813 \\
  & AFN+ & 0.8135 & 0.4386 & 0.7834 & 0.3798 & 0.7404 & 0.5945 & 0.9791 & 0.1824 & 0.9509 & 0.2583 \\
    & MaskNet  & 0.8137 & 0.4381 & 0.7835 & 0.3794
  & 0.7411 & 0.5935 & 0.9802 & 0.1783 & 0.9618 & 0.2372 \\
  & DCN-V2  & 0.8139 & 0.4382 & 0.7841 & 0.3775
  & 0.7443 & 0.5913 & 0.9823 & 0.1750 & 0.9624 & 0.2327 \\
  & FRNet& 0.8138 & 0.4383 & 0.7845  & 0.3774   & 0.7445 & 0.5909 & 0.9830 & 0.1607 & 0.9679 & 0.2278 \\ 
    & FinalMLP  & 0.8139 & 0.4380 & 0.7860 & 0.3772
  & 0.7443 & 0.5911 & 0.9832 & 0.1597 & 0.9670 & 0.2308 \\
  & \textbf{DELTA}        
  & \textbf{0.8147}$\star\star$ & \textbf{0.4374}$\star\star$   
  & \textbf{0.7878}$\star\star$ & \textbf{0.3768}
  & \textbf{0.7451}$\star$ & \textbf{0.5901}$\star\star$       
  & \textbf{0.9842}$\star\star$ & \textbf{0.1551}$\star\star$
  & \textbf{0.9690}$\star\star$ & \textbf{0.2191}$\star\star$
    \\ 
\bottomrule
 \end{tabular} 
 \caption{ Overall Performance of SOTA CTR models on Criteo, Avazu, Malware, Frappe, and MovieLens datasets. ``Lloss'' denotes the ``Logloss''.
 The one-tailed t-test shows that our performance advantage over previous SOTA methods is statistically significant over five datasets. ( $\star: p < 10^{-2}, \star\star: p < 10^{-4}$)
 }
 \label{tab1}
\end{table*}

\subsubsection{Experiment datasets}
We evaluate the proposed DELTA on the five challenging CTR datasets, including Criteo, Avazu, Malware, Frappe and Movielens. The detail of these datasets are summarized in the supplemental materials. Following~\cite{wang2022frnet,yang2020operation,cheng2020adaptive}, we randomly split instances by 8:1:1 unless specified for training, validation, and testing. 






\subsubsection{Evaluation metrics}

Our experiments employed two evaluation metrics: AUC (Area Under ROC) and Logloss (Cross-Entropy). It has been widely acknowledged in many works~\cite{chen2021enhancing,cheng2016wide,xu2021disentangled,lian2018xdeepfm} that an improvement of 0.001-level in AUC can lead to a significant increase in a company's revenue, particularly when the company has a large user base.

\subsubsection{Parameters settings}
We implement all models and experiments using pytorch~\cite{paszke2019pytorch}
. We set the batch size for all datasets to 4,096 and the learning rate to 0.0001 unless specified. The embedding size is 10 for Criteo, Avazu, and Malware and 20 for Frappe and MovieLens, respectively. For the first DNNs layers, we assign a [400,400,400] 3-layer DNN with a dropout rate equal to 0.5 following previous works~\cite{wang2022frnet}. For fair comparisons, we only use the same reported hyperparameters over five different datasets.

\subsubsection{Baselines}
We compare DELTA with the following eleven competitive methods, some of which are state-of-the-art models for CTR prediction.
Detailed descriptions of these methods are included in the related works and supplemental materials. We classify these methods into three types:

\begin{enumerate}
    \item Second-Order: \textbf{FM}~\cite{rendle2010fm}. It models both first-order and second-order feature interactions.
    \item High-Order: \textbf{NFM}~\cite{he2017neural}, \textbf{OPNN}~\cite{qu2018product}, \textbf{CIN}~\cite{lian2018xdeepfm}. They can model feature interactions higher than second-order.
    \item Ensemble: \textbf{DCN}~\cite{wang2017deep}, 
    \textbf{DeepFM}~\cite{guo2017deepfm}, \textbf{xDeepFM}~\cite{lian2018xdeepfm}, \textbf{AutoInt+}~\cite{song2019autoint}, \textbf{AFN+}~\cite{cheng2020adaptive}, 
    \textbf{DCN-V2}~\cite{wang2021dcn},  \textbf{MaskNet}~\cite{wang2021masknet}, \textbf{FRNet}~\cite{wang2022frnet}, \textbf{FinalMLP}~\cite{mao2023finalmlp}. These models adopt parallel or stacked structures to integrate different feature interaction methods. 
\end{enumerate}

\begin{figure*}[thb]
  \centering
  \includegraphics[width=0.9\linewidth]{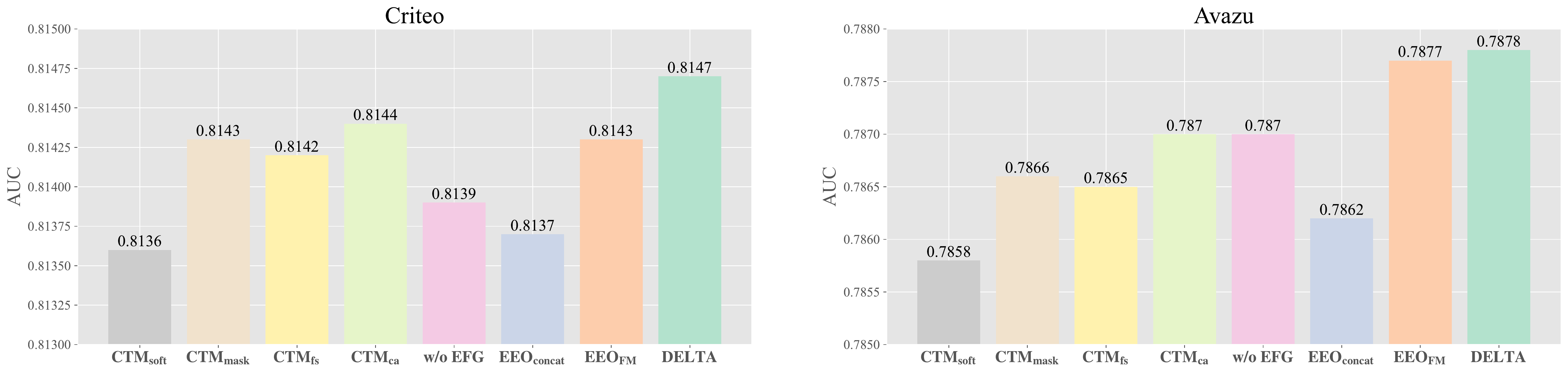}
  \caption{
  Performance comparison of DELTA module variants. 
  }
\label{ablation}
\end{figure*}

\subsection{Performance comparison (RQ1)}
We ran DELTA on every dataset 5 times and report the average results, then we conduct t-tests to compare DELTA with several strong baselines, and the results are summarized in Table~\ref{tab1}, from which we have the following key observations:

\begin{enumerate}
    \item In contrast to methods that use Deep Neural Networks to model interactions, FM and NFM underperform because they can only model second-order explicit feature interactions, which restricts their capabilities.
    \item Ensemble methods such as DCN-V2~\cite{wang2021dcn} and xDeepFM~\cite{lian2018xdeepfm} show robust performance across all datasets, demonstrating the effectiveness of combining implicit feature interactions and explicit feature interactions in the models.
    \item Dynamic embedding learning methods such as FRNet~\cite{wang2022frnet} and FinalMLP~\cite{mao2023finalmlp} show robust performance across all datasets. However,  MaskNet~\cite{wang2021masknet} lost its advantage on Avazu and Malware datasets, indicating that the utility of mask-based dynamic embedding learning methods might be data-dependent. 
    \item DELTA outperforms all other models in both AUC and Logloss on all five datasets. As shown in Table \ref{tab1}, the most significant improvement is observed on the Avazu dataset, where DELTA shows a relative improvement of 0.16\% over the second-best performing model.
    \item There is a trade-off between AUC and logloss across all models. When a model achieves the relatively highest AUC, it does not simultaneously achieve the relatively lowest log loss. For instance, AutoInt+ obtains the second-lowest Lloss while FinalMLP obtains the second-highest AUC on the Avazu dataset. This is due to the difference in optimization target between logloss and AUC, logloss focuses on the accuracy of prediction while AUC focuses on the order. In the real-world scenario, an increase in AUC will fundamentally boost the revenue\cite{zhang2021deep,cheng2016wide}.
\end{enumerate}
Overall, the results show the cutting-edge performance and generalization ability of DELTA on multiple datasets.

\subsection{Ablation study (RQ2)}

Here, we conduct experiments on Criteo and Avazu to prove that
the design of CTM, EFG, and EEO in DELTA plays an essential role in improving the performance of CTR prediction.
We compare DELTA with the following variants:
\begin{enumerate}
    \item $\mathbf{CTM_{soft}}$: Use the soft-attention~\cite{vaswani2017attention} to learn dynamic embeddings.
    \item $\mathbf{CTM_{mask}}$: Use the bit-wise mask introduced in MaskNet~\cite{wang2021masknet} to learn dynamic embeddings.
    \item $\mathbf{CTM_{fs}}$: Use the feature selection method introduced in FinalMLP~\cite{mao2023finalmlp} to learn dynamic embeddings.
    \item $\mathbf{CTM_{ca}}$: Use the context-aware method introduced in FRNet~\cite{wang2021masknet} to learn dynamic embeddings.
    \item $\mathbf{w/o}$ $\mathbf{EFG}$: DELTA without the EFG.
    \item $\mathbf{EEO_{concat}}$: Concat the output of EEO to the final layer of DELTA instead of applying it as an auxiliary task.
    \item $\mathbf{EEO_{FM}}$: Change the explicit interaction structure of EEO from cross-net to factorization machine.
\end{enumerate}

The ablation study results are presented in Figure \ref{ablation}. We can observe the particular design of every module in DELTA stably improves the performance.




\subsection{Hyper-parameter study (RQ3)}
\noindent \textbf{Consciousness bottleneck size of the CTM.}
We analyze the influence of the consciousness bottleneck's size in eq \ref{eq7} on the Criteo dataset in Table~\ref{topk}.
First, we empirically set the consciousness bottleneck's size to (39$\sim$19), we can observe that with the decreasing of the size, the AUC first increases and then decreases. Decreasing the bottleneck size will force the model to focus on the most important feature interaction, thus improving performance, while when the bottleneck is too small, vital information will inevitably be left out and derogate the performance of the model.
Then, we use the curriculum learning algorithm (``Curriculum'') to dynamically learn the bottleneck size, and the best performance shows its effectiveness.\\

\begin{table}[htb]
 \small
 \centering
  \begin{tabular}{c|c|c}
  \toprule
  \makebox[0.13\textwidth][c]{CTM size}       &  \makebox[0.13\textwidth][c]{AUC}           &  \makebox[0.13\textwidth][c]{Improvement}\\
\midrule
39 (Soft-attention)  & 0.8136 & / \\
34  & 0.8137 & 0.01\% \\
29  & 0.8141 & 0.05\% \\
24  & 0.8142 & 0.06\% \\
19  & 0.8139 & 0.03\% \\
14  & 0.8128 & -0.08\% \\
Curriculum  &\textbf{0.8144} & \textbf{0.08\%} \\

\bottomrule
  \end{tabular}
   \caption{Impact of consciousness bottleneck's size on the Criteo dataset. The best result learned by fixing $K$ (=24) is inferior to the dynamic $K$ (Curriculum learning). }
 \label{topk}
\end{table}

\noindent \textbf{Loss weight of the EEO.}
We investigate the impact of different loss weights of EEO on DELTA's performance from 0.0 to 0.7. As is shown in Table~\ref{tab3}, we can observe that the performance improvement is less influenced by the variance in the weight, and the performance is robust to the EEO weight in the range of (0.1$\sim$0.5). However, when the weight exceeded the threshold, it will face a performance degradation as the learning of embeddings is influenced by the coarse linear crossing to a great extent. During the training, we set the weight $\lambda = 0.5$ in the experiments.

\begin{table}[t]

 \small
 \centering
  \begin{tabular}{c|c|c}
  \toprule

 \makebox[0.13\textwidth][c]{EEO weight}     & \makebox[0.13\textwidth][c]{AUC}           & \makebox[0.13\textwidth][c]{Improvement}\\
 
\midrule
0.0   & 0.8134 & /       \\
0.1   & 0.8137 & 0.03\%  \\
0.3   & 0.8138 & 0.04\%  \\
0.5   & \textbf{0.8140} & \textbf{0.06\% } \\
0.7   & 0.8132 & -0.02\%  \\

\bottomrule
  \end{tabular}
   \caption{Comparison of different loss weights of the EEO branch on the Criteo dataset.}
 \label{tab3}
\end{table}

\section{Conclusion}
In this paper, we introduce a new CTR framework called DELTA that incorporates human conscious processing.
We investigate a conscious truncation module (CTM), which leverages curriculum learning to learn adaptive truncation on attention weights to select the most critical feature under different contexts to learn dynamic feature representations. 
We further improve the learning of embedding representations by proposing a explicit embedding optimization (EEO), which independently propagates gradient from the loss layer to the embedding layer to explicitly enhance the crucial embeddings. 
Note that the simple yet effective EEO can be simply removed and requires no extra cost during inference. 
The experiment results on five real-world CTR datasets demonstrate that our DELTA outperforms the state-of-the-art methods.

\bibliography{aaai24.bib}

\end{document}